# Standing Together
## for
## Reproducibility in Large-Scale Computing

## Report on reproducibility@XSEDE
**An XSEDE14 Workshop**
**July 14, 2014**
**Atlanta, GA**

Developed collaboratively by the reproducibility@XSEDE workshop participants[1]

Principal Editors:
Doug James, Nancy Wilkins-Diehr, Victoria Stodden, Dirk Colbry, and Carlos Rosales

Finalized 17 Dec 2014

**Abstract.** *This is the final report on reproducibility@xsede, a one-day workshop held in conjunction with XSEDE14, the annual conference of the Extreme Science and Engineering Discovery Environment (XSEDE). The workshop's discussion-oriented agenda focused on reproducibility in large-scale computational research. Two important themes capture the spirit of the workshop submissions and discussions: (1) organizational stakeholders, especially supercomputer centers, are in a unique position to promote, enable, and support reproducible research; and (2) individual researchers should conduct each experiment <u>as though</u> someone will replicate that experiment. Participants documented numerous issues, questions, technologies, practices, and potentially promising initiatives emerging from the discussion, but also highlighted four areas of particular interest to XSEDE: (1) documentation and training that promotes reproducible research; (2) system-level tools that provide build- and run-time information at the level of the individual job; (3) the need to model best practices in research collaborations involving XSEDE staff; and (4) continued work on gateways and related technologies.  In addition, an intriguing question emerged from the day's interactions: would there be value in establishing an annual award for excellence in reproducible research?*

## Overview

The reproducibility@XSEDE workshop [26] took place on Monday, July 14, 2014 in Atlanta, GA in conjunction with XSEDE14 [42], the annual conference of the Extreme Science and Engineering Discovery Environment (XSEDE)  [41]. The workshop featured an interactive, open-ended, discussion-oriented agenda that focused on reproducibility in large-scale computational research. Consistent with the overall XSEDE14 conference theme, the organizers sought to engage participants from a broad range of backgrounds, including those whose computational interests extend beyond traditional modeling and simulation as well as decision-makers and other stakeholders whose work informs and determines the direction of computation-enabled research. The organizers intended an agenda that extended beyond best practices within research groups, focusing instead on community needs and priorities, especially (1) delivering and maintaining a robust cyberinfrastucture that enables reproducible research at scale; and (2) promoting a culture of reproducibility within the broad community of stakeholders.

---

[1] See Appendix B below.

The event built on the 2009 Yale Data and Code Sharing Roundtable [36], which culminated in a declaration "demanding a resolution to the credibility crisis from the lack of reproducible research in computational science." The organizers planned the XSEDE14 workshop as a response to the challenges issued in the Yale 2009 declaration.

In the months leading up to the workshop, the committee sought short, informal reflections in the spirit of the Yale 2009 submissions [36] that: (1) connected the Yale 2009 Declaration to the needs, priorities, constraints, and circumstances of communities like XSEDE; (2) reflected a strong, informed point of view; and (3) included specific, well-reasoned recommendations. A total of thirty-five authors and co-authors contributed twelve submitted position papers. Accepted submissions, which by design did not undergo formal peer review, appear in the bibliography below and are available from the workshop home page [26].

A total of 26 participants from inside and outside XSEDE, listed in Appendix B, gathered in Atlanta for the workshop itself. Victoria Stodden led the initial session (Session 0 below: "Setting the Tone"). Her featured presentation "Reproducibility in Computational Science: Why? What? and How?" framed the workshop agenda by connecting the current needs of the large-scale computing community to initiatives and discussions in the recent past, especially the Yale 2009 event [33]. She related the 2009 aspirations to the recommendations in the 2012 ICERM workshop report [35], building a context for the current workshop discussion.

Stodden then moderated the first of four open discussions. Dirk Colbry (Session 1 below: "Where Do We Stand?"), in a session that began with some thought-provoking lightning talks, then guided the group through a discussion intended to explore the state of the practice in large-scale computation. Lorena Barba (Session 2 below: "What Can We Learn?") broadened the focus in the next session: another round of energetic lightning talks provoked a lively discussion regarding the ways we can learn from and leverage the experience of the broader community. The workshop concluded with "What Comes Next?" (Session 3 below), moderated by Nancy Wilkins-Diehr: a wide-open, no-limits brainstorming session regarding potentially promising activities, initiatives, and priorities. Following the event in Atlanta, the participants collaborated to produce this final report.

**Observations and Conclusions**

Reproducibility means many things to many people. In this sense the workshop participants were representative of the larger community: both the submissions and discussions represent a wide range of perspectives and priorities. Two quotes, however, resonated among the participants and capture the spirit of the workshop:

- Mark Fahey expressed a view common among participants when he said that "**supercomputing centers [and other stakeholder organizations and communities] need to step up and accept some responsibility**" for empowering individuals and research groups to conduct reproducible research. This is more than a generic statement about organizational responsibilities; it reflects the reality of the unique relationships that connect supercomputing end-users and resource providers.

- Similarly, Lorena Barba struck a chord when she challenged us to "**conduct our research *as though* someone will repeat our experiments**." This is appropriate whether or not we[2] or anyone else will ever do so in the literal sense, and remains so even if literal replication isn't even feasible. Her words captured well the idea that reproducibility is more than a characteristic of an experiment: reproducibility is an attitude modeled by those conducting research, and a mindset, culture, expectation established and cultivated by our peers and all those to whom we are accountable.

The worldview captured by these quotes informed and affected the entire day's discussion, much of which focused on two areas in which the needs and realities of large-scale computing are unique:

- **Scale.** Throughout the day, participants found themselves energized by Barba's challenge to conduct research *as though* someone will repeat our experiments. At least for this one day, this was a higher priority than discussing the technical and programmatic challenges of actually repeating experiments at scale. One participant put it bluntly: "we can't virtualize Blue Waters."
- **Control and Visibility.** One important but often neglected difference between the desktop and supercomputer relates to control over and visibility into the end-user's environment. Hardware and software choices, updates, and configuration changes are largely in the hands of others. Moreover, it can be difficult for end-users to gain insight into the voluminous details and assess the potential or actual impact of changes to a system's configuration. The result is that users often feel that the ground is shifting under their feet: it is essentially certain that a large cluster's configuration is different this week than it was a week ago, and seemingly minor (often invisible) changes can have unintended and often undetected effects on an experiment.

Acknowledging this reality, the workshop participants see great value in system-level tools and initiatives that mitigate this problem by returning some degree of control back to the researcher and giving end-users greater visibility into a system's configuration. Such tools and initiatives include a wide range of potentially high impact enhancements to the cyberinfrastructure, in particular (1) scientific gateways and related tools that simplify, streamline and automate domain-specific workflows; and (2) behind-the-scenes system-level services that track, audit, analyze and summarize user activity at the level of the individual job.

Despite the unique challenges associated with large-scale computing, the discussions suggest that participants clearly understand that many of our realities, needs, and priorities are closely aligned with the broader community engaged in thinking about reproducible research. We acknowledge the hard-won experience of this community, and seek to leverage that experience in the service of those engaged in large-scale computation. It is not a stretch to suggest there is a large degree of consensus within the broader community regarding the nature of our common experience and its implications. Most would agree, for example, that there is an acute need for discipline and rigor that is outpacing common practice in the trenches. Given the high stakes impact and broad scope of modern computational research, we would also agree on the importance and value of such discipline and rigor. Finally, we recognize that there is little controversy regarding best practices for individuals and small research teams: e.g. change management and automation [28]. Thus, we encourage and advocate community investment in promoting and teaching these practices. In general, however, we caution against the need to re-invent the wheel by developing from scratch high quality resources that are already available elsewhere.

---

[2] Here we emphasize a lesson many of us have learned the hard way: replicating *our own* results is often as challenging as reproducing somebody else's.

Workshop participants from within the XSEDE community were especially interested in identifying opportunities and initiatives worth promoting within XSEDE. Recognizing that XSEDE is in an interesting and perhaps unique position to pilot ideas for the broader community of large-scale computing stakeholders, we organized our concluding brainstorming activity (session 3 below) around the XSEDE organizational structure, and documented issues, questions, technologies, practices, and actions [27] that may warrant further investigation and continued discussion. In particular, several potentially promising organizational priorities seemed to resonate among XSEDE participants:

- Actively seek opportunities to develop and structure documentation and training materials in ways that encourage and enable reproducible research. When possible, promote and leverage existing resources, e.g. training activities like those pioneered by the Software Carpentry Initiative [39] and literature like *Ten Simple Rules for Reproducible Computational Research* [28].
- Encourage XSEDE Service Provider (SP) organizations to (1) evaluate XALT (see [5], [1]) and other system-level tools that promote reproducible research; and (2) share their recommendations with XSEDE Campus Champion sites. XALT automatically collects, records, analyzes, and summarizes detailed build- and run-time software usage data (compilers, libraries, and version information) for all users' jobs; such information is likely to be valuable to stakeholders at all levels, including decision-makers, support staff, and end-users themselves. This rapidly maturing tool, currently deployed on production systems at the National Institute for Computational Sciences (NICS) and the Texas Advanced Computing Center (TACC), is available for beta testing.
- Model best practices in formal research collaborations between XSEDE staff members and end-user research groups.[3] While this certainly means conducting research activities in a reproducible way and mentoring research groups in process-related matters as appropriate, it could mean more: e.g. addressing reproducibility in work plans and final reports; providing training opportunities to staff and end users alike; etc.
- Continue, perhaps expand, XSEDE's extensive efforts to promote, develop, deploy, support, maintain, and improve gateways and related technologies [38] that make possible user-friendly workflows, automated processes, and repeatable results for users whose priority, experience and focus is domain-specific research rather than software process.

Finally, we offer a somewhat unconventional idea intended as a response to those tempted to throw up their hands in frustration at the enormity of it all. At one point during the day someone offered an aside, almost to herself: would there be value in establishing an annual award for excellence in reproducible research? The room got a little quieter. What would such an award look like? We don't know -- we didn't discuss the matter. But some of us left the workshop wondering whether such an award might focus the broader conversation (and raise its visibility) in ways that might otherwise be difficult to achieve.

---

[3] See additional references to the Extended Collaborative Support Service (ECSS) elsewhere in this report.

**Summary of Individual Workshop Sessions**

Slides for all talks are available on the agenda page of the workshop website [26].

***Session 0: Reproducibility in Large-Scale Computing: Setting the Tone (Victoria Stodden)***

The workshop opened with Victoria Stodden's very motivating introductory talk. Victoria observed that we stand on the shoulders of giants, and then set the stage by summarizing recent news from both the popular press and the scientific community demonstrating the need for and importance of reproducible research.

Any discussion of reproducibility must at least acknowledge different understandings of the concept. One plausible definition has *repeatability* as its focus: can we replicate an experiment, at least in principle? Another approach focuses on independent *verification*: can we reach the same conclusions within appropriate thresholds using independent implementations of the relevant (or alternative) algorithms? Workshop participants also highlighted other distinctions that can be important. It is often useful, for example, to distinguish among different types of data: in particular data as an input to a computational model or simulation versus data as an output (this includes traditional output files, but also describes images and other visualizations). Similarly it is often important to distinguish source code and executables from more traditional data types; using the term "data" too broadly is often confusing at best.

One thing is clear: it is not enough to share only traditional research publications. Reproducibility requires access to codes, workflows, environments, and data. In 1996 the Human Genome Project developed the Bermuda Principles [12], documenting community consensus for reproducibility in genome decoding and managing the results (e.g. requiring data to be released in publicly accessible databases within 24 hours). In 2009, the Yale Roundtable [36] gathered a diverse group from chemistry, biology, astronomy, computer science, mathematics, law, medicine and more who collaboratively assembled nine "dream" goals to move computation in the direction of reproducible research.

We have matured since 2009. We can document the context of simulations, including the input, the code, and (at least to some extent) the system environment. We can save code and data in permanent repositories, and automate workflows in unprecedented ways. In some cases we can operate in virtual, completely reproducible worlds. We have not, however, found mechanisms for sustaining much of this activity. To cite one simple example: how do we fund access to and maintenance of data after the original research grant expires?

The collaboratively authored workshop report [35] from a December 2012 workshop held at Brown University's Institute for Computational and Experimental Research in Mathematics (ICERM) included fifteen best practices for preparing publications to support reproducibility:

1. A precise statement of assertions to be made in the paper.
2. A statement of the computational approach, and why it constitutes a rigorous test of the hypothesized assertions.
3. Complete statements of, or references to, every algorithm employed.
4. Salient details of auxiliary software (both research and commercial software) used in the computation.

5. Salient details of the test environment, including hardware, system software and the number of processors utilized.
6. Salient details of data reduction and statistical analysis methods.
7. Discussion of the adequacy of parameters such as precision level and grid resolution.
8. Full statement (or at least a valid summary) of experimental results.
9. Verification and validation tests performed by the author(s).
10. Availability of computer code, input data and output data, with some reasonable level of documentation.
11. Curation: where are code and data available? With what expected persistence and longevity? Is there a site for future updates, e.g. a version control repository of the code base?
12. Instructions for repeating computational experiments described in the paper.
13. Terms of use and licensing. Ideally code and data "default to open", i.e. a permissive re-use license, if nothing opposes it.
14. Avenues of exploration examined throughout development, including information about negative findings.
15. Proper citation of all code and data used, including that generated by the authors.

Workshop participants also discussed taking the step of assigning Document Object Identifiers (DOIs) to data, code, and workflows that are associated with research articles that have themselves been assigned a DOI. Several argued that data, code, and other digital scholarly objects should be released under permissive open licenses such as the MIT License or Modified BSD License that permit researchers to share and re-use these objects [32].

Participants lobbied for permanent persistence of digital scholarly objects, although some noted the current standard seems to be 10 years. The discussion also touched on the need to work closely with campus libraries, particularly on matters related to archiving code, encouraging citation standards, and assigning DOIs.

Other thoughts, recommendations and potential action items emerged from the discussion:

- Reproducibility could be offered as an XSEDE service, and results could be certified under specific standards for reproducibility.
- Specialized workshops focusing on technologies like Cactus [3] and XSEDE may be worth considering. Some Cactus thorns (application modules) focus on large-scale infrastructure.
- What is the best way to establish and maintain clear and robust connections between published results and the associated artifacts (codes, data, workflows) associated with those results? Are there tools, practices, or mechanisms that will enable the research community to move easily between a publication and the related code/data ?
- Deploy a system for tracking build- and run-time information (machine state, input parameters, data input versions, precise functions called, etc.) as part of the XSEDE framework. See discussions about work in this direction (e.g. XALT [5]) elsewhere in this report.
- Investigate questions related to meta-data: what kind of information stored with datasets will encourage and promote re-purposing of the data?
- Practices that aim for reproducibility produce better research, even if the work is never actually reproduced (e.g. due to cost) so encouraging these practices for their own sake is worthwhile.

***Session 1: Reproducibility in Large-Scale Computing: Where Do We Stand? (Dirk Colbry)***

The session began with lightning talks by Mark Fahey, Rob Kooper, Justin Shi, and Nancy Wilkins-Diehr, summarized in Appendix A below. Shi added later that "[m]any modeling/programming flaws will only show up in large-scale tests. We can apply [this] idea to the current main stream programming models. The issues become obvious...The fundamental solution to improve the reproducibility of any program is to reduce dependencies. The current mainstream explicit parallel programming models should be considered counter-productive toward this goal. " His written submission [31] and lightning talk highlight his proposed alternative.

After the lighting talks the group engaged in a lively general discussion about the current state of reproducibility. Much of this early discussion compared and contrasted different definitions of reproducibility, including but not limited to algorithmic, computational, bitwise, statistical, experimental, and semantic[4] reproducibility. Although the participants steered away from splitting hairs, they certainly agreed that there are many different types and levels of reproducibility. As a minimum, this means that those engaged in discussions on reproducibility should understand that their peers may be working from definitions different from their own.

One general consensus was that bitwise reproducibility is often an unrealistic expectation; machines, programming environments, and software libraries all change. Instead it is usually enough to replicate results within appropriate error bounds. In an ideal world, one might expect that researchers would be able to reproduce the results from just a description of an algorithm. However, this high level of reproducibility is not reasonable in many cases. Generating datasets and new code bases is expensive. The time and effort to reproduce a scientific result from just an algorithm is just not practical in many scientific disciplines. Instead, other levels of reproducibility may be more appropriate.

Social incentives in modern academic culture reward new science, but there are not yet enough comparable mechanisms to encourage the level of process discipline that makes possible appropriate levels of reproducibility. The group agreed that it is important to adjust the academic culture to better reward reproducibility. Researchers should encourage replication and validation of their own results by making it as easy as possible for other scientists to do so. For this reason, scientists should capture all of the characteristics of their experiments with as much detail as necessary. These details include but are not limited to: full access to documentation of source code, input data, computing environment, and published results. Researchers should be doing this anyway to ensure continuity in their own research programs when they move to new computational platforms or transition between students in their own labs.

The discussion then led into best practices for researchers today. There are many existing technologies (including version control, unit testing, standard libraries and build systems) that can make life easier for those in the computational trenches. However, not everyone knows (or believes) this. Thus one key to promoting reproducibility is educating the research community on best practices and tools. Participants mentioned existing programs such as HPC University [11], the XSEDE tutorials, and Software Carpentry[5] (and soon Data Carpentry) [39] as excellent models for training our next generation of scientists.

---

[4] On semantic reproducibility, see summary of Laurence Loewe's contribution in Session 2 below.
[5] Special thanks to Greg Wilson for responding graciously and rapidly to feedback from a brief side discussion. One workshop participant expressed mild frustration that the phrase "boot camp" might be off-putting to some who would otherwise be supportive of the Software Carpentry initiative. Greg immediately changed the terminology.

There is also an important role for resource providers in helping researchers with reproducibility. Computing centers need to provide detailed descriptions about hardware, operating system, compilers and library versions used in individual experiments. Ideally this information is stored so it can be tracked as it changes over time. XALT [5] is now a maturing project that has moved beyond its prototype/predecessors ALTD and Lariat in significant ways. Tools like this offer great promise, but the ideas and the tools themselves are still new to this community. Virtual Machines (VMs) and related technologies like Docker [13] and Vagrant [9] are poised to revolutionize the world of reproducibility, but cannot yet operate at large-scale (high core counts).

Publishers and funding agencies also have a responsibility to ensure that reproducibility-related products and activities are part of the review process. The academic landscape of today already includes increased attention to academic rewards for software and data, reuse, computation, data transfer. These rewards still need to propagate to the tenure review process in order to gain widespread acceptance.

As the session concluded, the general sense seemed to be this: although we are making progress, achieving long-term reproducibility is still a difficult problem. Existing challenges include obsolete operating systems, software availability, and files that are no longer readable due to obsolete, proprietary formats.

### Session 2: Reproducibility in Large-Scale Computing: What Can We Learn? (Lorena Barba)

The session began with lightning talks by Rafael Silva, Kyo Lee, Ralph Roskies, and Doug James, summarized in Appendix A below. Discussion participants struggled to define reproducibility in the context of large-scale computing. Should reproducibility mean that the original code, data and computing environment is effectively packaged and preserved to be re-run at any point in the future? Or does reproducibility not necessarily imply preservation? One could argue, for example, that reproducibility involves the ability to faithfully model the semantics of the computational experiment and replicate that experiment at any future date on whatever technology may exist. The latter is akin to the traditional interpretation of reproducibility in the realm of experimental sciences, while the former represents a newer, data-driven and therefore more restrictive view.

The general consensus seemed to be that code should be well documented and open to scrutiny in order to have any claim that its results are reproducible. Practices that lead to well documented code and research work included automated processes and workflows; traceability; use of change management tools; as well as clear information regarding data provenance, format and context (metadata).

Several workshop participants challenged us to think about reproducibility in new ways. Laurence Loewe, for example, expressed a special interest in semantic reproducibility, defined as the ability to transfer meaning between the encoder of a symbolic message and the decoder of that message, without changing the meaning. In his words, "systems for semantic reproducibility have been used for a long time; we usually call them 'languages' and take for granted that they work...Why do professional programmers insist on good software practices and well written code with reasonable function names? Because they know the cost of having to decipher the hidden meaning in other people's code. Thus learning a new code base is in a way not dissimilar form learning a new language. This makes semantic reproducibility a hard problem that requires experimenting with innovative approaches. The BEST Names concept [17], currently in the earliest conceptual phase, might be one such approach."

Participants acknowledged that operating at scale (large core counts) certainly introduces new complications and issues: there are different tools and environmental challenges associated with reproducibility at large scale. Replicating large-scale experiments may be impractical for any number of reasons, including cost and availability of suitable computing resources. Conducting (and documenting) all experiments *as though* someone will replicate the results can increase confidence that the results are trustworthy even if the experiment itself is never actually repeated.

The group also discussed the value of reproducible processes within a research group or (especially) a geographically dispersed collaboration. Training the next generation in reproducibility certainly has long-term benefits for the community as a whole. But there are short-term benefits as well: when your students graduate and your post-docs move on, you will be very glad to know that their replacements will be able to pick up where their predecessors left off!

Finally, the participants turned again to the question of incentives for reproducible work. Our current research environment provides insufficient rewards to researchers who work in a reproducible manner. The widespread view (and arguably the reality) is that funding depends solely on citation levels and publication numbers; some see little value in time and effort spent on process-related priorities. The counter-arguments are clear, however: reproducible techniques lead to higher quality results and the trust and respect of the community. The community must provide incentives and encourage the products and processes that are the heart of reproducible research. Such incentives need not be tied to funding in the narrow sense. One could imagine, for example, a culture in which researchers are more likely to cite reproducible work. How do we begin to move the community in this direction? While we know of no shortcut or magic bullet, the group seemed sympathetic to any and all ways to raise visibility and awareness. Thus the idea of an annual prize for excellence in reproducibility struck many as an intriguing one.

### Session 3: Reproducibility in Large-Scale Computing: What Comes Next? (Nancy Wilkins-Diehr)

Session 3 included no lightning talks, but was an opportunity for workshop participants to reflect on discussions throughout the day and put forth ideas for the future.

The organizers had several preliminary outcomes in mind when planning this workshop. The first was simply to highlight the issue, begin a discussion within XSEDE, and increase awareness in the XSEDE community about computational reproducibility and the special challenges of conducting reproducible research at scale. Many in XSEDE (both staff members and users) are not aware of the critical credibility issues facing computational science today.

The second goal was to understand what it might take to promote a culture of reproducibility among stakeholder organizations and their users. In particular, what can XSEDE do in the way it conducts its business that supports users in their quest to conduct reproducible computational research? The third goal was to communicate to XSEDE leadership the ideas generated at the workshop.

This section of the report captures all ideas discussed during this session, organized by XSEDE area. Other large scale computing programs may recognize aspects of their own organizations and so may be able to map some these ideas to their own needs. Workshop attendees used Mozilla's Etherpad to collaborate in real time to produce a record of the discussion and individual, author-attributed reactions to that discussion. A read-only record of these notes is available at [27] with authors listed at the end of the pad.

The group began by noting on-going activities and practices among vendors, centers and users that already support reproducibility. Vendors check for reproducible results when designing new chips. Centers run acceptance tests on new platforms that include application suites. While performance and reliability are generally the criteria here, comprehensive regression tests ideally ensure that codes get the same results on a new system. In addition, many centers run such tests regularly after hardware or software upgrades to validate results systematically. Users typically port codes from one system to another every few years and will run tests to check results. Similarly when users add code to incorporate new science they will also typically check for consistent results. Many of these activities, however, are ad hoc.

Lack of information about the user environment (operating system, environment variables, dynamic libraries, etc.) when a job runs was highlighted as a challenge for those doing large-scale computing. Participants discussed several tools in use at centers to capture runtime information for users, especially XALT [5], a joint NICS-TACC project based on TACC's Lariat and NICS' ALTD prototypes. Mark Fahey discussed these tools in a lightning talk during session 1 (above).

This suggestion led to several ideas for discussion amongst XSEDE service providers (SPs). The Service Provider Forum of XSEDE may want to evaluate packages like XALT [5]. Packages viewed as valuable could be then recommended to Campus Champion sites as well. Regression tests used after system upgrades could be shared among SPF members. One workshop participant also recommended Nix [23], a package management system for Linux and Unix intended to make "package management reliable and reproducible". A second participant suggested developing a system tool that would allow users and administrators to record login sessions and command histories.

In the area of XSEDE Training, Education and Outreach (TEOS) and User Services, several ideas were put forth to support reproducible research. Develop, contribute and use reproducibility training content to the Mozilla Foundation's Software Carpentry program [39]. Encourage participation among XSEDE's Campus Champions, users and students in reproducibility training. Consider a reproducibility challenge, prize or technical paper track as part of the annual XSEDE conference. Include reproducibility content in the XSEDE TEOS group's work on curriculum development, including a possible certification in reproducibility. Schedule regular "hangouts on air" with XSEDE staff and users to discuss best practices and technologies that support reproducibility.

Participants also suggested that XSEDE produce documentation at various levels (beginner, intermediate, expert) that instruct and demonstrate XSEDE's commitment to reproducibility. This would be particularly valuable for new projects that could more easily incorporate "best practices" or "best tools" recommendations. Attendees thought such documentation could be valuable for early career computational scientists and those studying computational science. Testing support tools named at the workshop include googletest [8] (see also googlemock [7]); Mocha [20]; and Jenkins [15], an application that monitors the execution of repeated codes.

There were also suggestions related to XSEDE Allocations. Currently users are required to show scaling and performance information in their requests; XSEDE could add reproducibility as an additional requirement. This, however, places additional burdens on those submitting allocation requests as well as reviewers. Additionally, reproducibility claims can be difficult to prove. One workshop participant shared a link to a thoughtful examination of research programs modeling excellence in reproducibility [10].

Participants mentioned services that promote reproducibility by tying data to publications and storing citable data. The DataVerse Network Project [16] supports datasets typically under 1 TB. An institutional implementation for Purdue researchers is described at [24]. XSEDE may want to investigate how such to provide such a service for larger datasets.

XSEDE's Extended Collaborative Support Service (ECSS) program [41] makes possible distributed collaborations between a PI's research group and XSEDE staff members. ECSS staff members make code improvements to optimize performance, but also must be sure that a code's results don't change as a result of the optimization. One recommendation from the workshop is to explicitly add activities that support reproducibility into work plans that staff and PIs develop jointly for ECSS projects.

ECSS collaborations require using and promoting sound programming practices; e.g. use of version control software. Currently XSEDE does not host its own installation of packages like CVS, SVN and Git. There have been discussions among ECSS management and staff about whether XSEDE hosting of such packages is necessary. Many user groups use public installations (open source requirement) or maintain their own installations. One participant suggested coupling services to GitHub URLs to promote the use of version control, to make things easier for people who have their code in a public version-control repository.

ECSS staff may also wish to address reproducibility-related issues in their work with community codes. At some XSEDE service provider sites, community codes are part of regression tests, but it would be useful to have a formal look at reproducibility as users move across platforms and as platforms are upgraded. XALT could help with prioritization by tracking usage of these community codes. The ECSS Community Codes area could then start with an evaluation of heavily used codes in cooperation with code developers.

ECSS has also recently added expertise in the workflow area. This group could work to describe how various workflow packages do (or do not) make it easy to automate processes, track provenance, and preserve metadata related to computational experiments. In addition, XSEDE now conducts staff training workshops for ECSS staff; these could include topics related to reproducibility and responsible programming.

Discussions then moved beyond specific XSEDE organizational areas. An XSEDE "reproducibility czar" could coordinate activities across XSEDE functional areas. The discussion then turned to data management plans as a mechanism for connecting publications to both traditional data and software. XSEDE-hosted projects as reproducibility examples (with documentation and quality control) could be potentially valuable if they belonged to established leaders in specific fields. Participants also suggested following and promoting work done by others, for example the first International Workshop on Reproducibility in Parallel Computing, held in Portugal in August 2014 [25].

The group engaged in this brainstorming session committed to recording all ideas and every participant's input [27], and knowing that any given item might warrant further investigation and continued discussion within or outside of XSEDE. XSEDE participants did, however, elect to highlight four high-interest areas in the Observations and Conclusions section at the top of this report.

## Appendix A: Summary of Lightning Talks

Those who submitted position papers to the workshop received as a professional courtesy a five-minute speaking slot for one author or a representative designated by the authors. Most speakers summarized their written submission, and this report's bibliography includes all submitted papers. The organizing committee, however, did give speakers the freedom to use their allotted time in any reasonable and appropriate way. Follow the hyperlinks below or on the agenda page at [26] for lightning talk slides.

Mark Fahey. A part of reproducibility is having complete documentation about how a simulation was run so another researcher could replicate the experiment. This documentation should include the code, its version, and hopefully a manual, the inputs to the code, the computing environment (hardware and software specifications), how the code was built (compiler options and libraries), how the code was run (runtime environment settings, number of processors, etc.), and other instructions necessary to run the code and duplicate the experiment. And of course the output data, any analytics derived from the data, and where the results are published are also components of the documentation. In many publications, the code, the compiler, and the machine are specified, but the rest (compiler options, libraries, versions of each component, programming and runtime environments) are left out. It has even been suggested that the researcher should include a virtual machine replicating the environment that can reproduce the experiment. Some of these aspects the researcher could reasonably add into their documentation, but some are not so reasonable like expecting a researcher to document the computing environment in which he ran when the computing environment may be a world-class supercomputer with customized operating systems and programming environments. Further, it is impossible to expect a virtual machine can replicate this environment, and worse to expect the researcher to construct it. Therefore, the computing centers themselves have to provide the missing pieces of documentation and it is actually quite feasible. Centers can automatically collect information for each build and experiment, and this information can be made available to the researcher for publications if desired. There are already two prototypes (ALTD and Lariat) at a variety of computing centers around the world that collect a good portion of this information, and the NSF-funded XALT follow-on [5] is now available for beta testing. Computing centers can collect this information and they should be expected to do so and give users access to it, and then the documentation problem is not as big a problem anymore.

Doug James. There is more to reproducibility than re-enactment. Rather than run your code, I may want to implement your algorithm in my own way; turning two different cranks may be a better way to gain confidence in our results than turning the same crank twice. Rather than duplicate your hardware and software environment, I may prefer to run your experiment in my own environment in pursuit of results that are independent of the choice of environment. And rather than expecting bitwise-identical results, I may prefer to honor the realities of floating point arithmetic, asynchronous execution, compiler optimizations, and problems that are sensitive to their inputs; it may be that I prefer to *understand* the differences between our results rather than try to eliminate them. These priorities do in fact reflect my own preferences. So you will not be surprised that (1) I am in favor of organizational priorities that promote the basics over exotic initiatives, and (2) I hope to see our precious resources brought to bear on promoting the value of common sense fundamentals: automation, change management, traceability, and quality documentation. I would also like to see a world in which solicitations at least encourage us to think about and comment on issues of reproducibility: perhaps a sixth item in the NSF Grant Proposal

Guide II.C.2.j (Data Management Plan) [21] that reads something like "... mechanisms and processes that enhance opportunities for reproducibility."

Rob Kooper. We attempt to address a number of the issues facing scientific reproducibility within the information age, specifically those surrounding long tail scientific data. Towards this we propose two services, the DAP and DTS, which each address a handful of the considerations being faced and provide services by which to allow users to better search, access, and use uncurated data collections while simultaneously serving as a means of preserving, advertising, and crediting the code used to populate these services. Lastly, we attempt to do this in a manner that also serves the general public towards the sustainability of the two services.

Kyo Lee. Climate science is Big Data Science. Big data in climate science have some unique characteristics that we need to consider regarding reproducibility of scientific results. Both observational and model datasets are extremely distributed, and this distributed nature makes data access and analyses hard for users especially those with limited computation resources. As a result, quickly and continuously updated database for studying climate science makes it hard to replicate previous studies. As one of potential solutions for the reproducibility issue in climate science, the National Aeronautics and Space Administration's Jet Propulsion Laboratory developed open source software, Regional Climate Model Evaluation System (RCMES). RCMES, which is a combination of database and toolkits, facilitates reproduction of previous model evaluation studies via easy access to observations focusing on remote sensing datasets and open source toolkits with user-friendly interface. The database is largely scalable to store observational data from a variety of sources in a common format. The toolkit part includes libraries to process data, to calculate model evaluation metrics and to visualize calculation.

Ralph Roskies. OLIVE is a system that freezes and precisely reproduces the environment necessary to execute software long after its creation. It uses virtual machine (VM) technology to encapsulate legacy software, complete with all its software dependencies. This legacy world can be completely closed-source: there is no requirement for availability of source code, nor a requirement for recompilation or relinking. The entire VM is streamed over the Internet from a web server, much as video is streamed today.

Justin Shi. Dependencies in software and hardware are the biggest enemies in reproducibility of scientific computational research. Fundamental research is needed to reexamine the current programming paradigms and models for extreme scale computing in order to minimize dependencies without compromising performance and reliability. A decoupling implicit data parallel <key, value> tuple space paradigm and a tuple switching network (statistic multiplexed computing or SMC) were proposed to address the reproducibility and scalability issues at the same time. Small scale experiments showed the feasibility of finding the completion time equilibrium in parallel computing tasks that outperformed explicit parallel counterpart. The inductive nature of SMC architecture promises to overcome perceived architectural inflection points and deliver better performance and better reliability than explicit parallel programs in extreme scales with minimal software and hardware dependencies.

Rafael Ferreira da Silva. Reproducible research in scientific workflows is often addressed by tracking the provenance of the produced results. While this approach allows inspecting intermediate and final results, improves understanding, and permits replaying a workflow execution, it does not ensure that the computational environment is available for subsequent executions to reproduce the experiment. In this work, we propose describing the resources involved in the execution of an experiment using a set of semantic vocabularies, so as to conserve the computational environment. We define a process for documenting the workflow application, management system, and their dependencies based on 4 domain ontologies. We then use those descriptions to define the infrastructure specification. This specification can then be used to derive the set of instructions that can be executed to obtain a new equivalent infrastructure. Experimental results show that our approach can reproduce an equivalent execution environment of a predefined virtual machine image on an academic and a public cloud platform.

Nancy Wilkins-Diehr. As noted by other speakers at the workshop, the lack of reproducibility in science is being increasingly highlighted in the news media. Concurrently, computational science and simulations are being increasingly relied upon to serve as surrogates for physical experiments. This role means that computational scientists must also conduct their work in a manner that supports reproducibility. Many researchers rely on software frameworks to conduct their computational simulations. Examples include IPython notebooks, Galaxy [6], Vistrails [37], and nanoHUB [22]. If instrumented correctly, these frameworks, also known as science gateways can also be used in support of reproducible science. The question posed to workshop attendees was what do gateway developers need to consider to enable this.

## Appendix B: Contributors

Allen, Gabrielle (E);  Bailey, David (W);  Barba, Lorena (EIC);  Borwein, Jonathan (W);  Cazes, John (E); Cho, Kym Won (E);  Colbry, Dirk (EIWF);  Corcho, Oscar (W);  Deelman, Ewa (W);  Dietze, Michael (W); Fahey, Mark (EIWF);  Fonseca, Jim (E);  Gilbert, Benjamin (W);  Harkes, Jan (W);  Hovig, Elvind (C); Hwang, Lorraine (E);  James, Doug (EIWCF);  Katz, Dan (EC);  Kay, Sophie (C);  Keele, Seth (W);  Koesterke, Lars (E);  Kooper, Rob (EIW);  Kumar, Praveen (W);  Lee, Jong (W);  Lee, Kyo (EIF);  Lindsey, Susan (EF); Linke, Erika (W);  Loewe, Laurence (EWF);  Marciano, Richard (W);  Marini, Luigi (W);  Mattman, Chris (W);  Mattson, Dave (W);  McHenry, Kenton (W);  McLay, Robert (W);  Miguez, Sheila (W);  Minsker, Barbara (W);  Patel, Pragnesh (E);  Perez-Hernandez, Maria (W);  Pouchard, Line (E);  Rosales, Carlos (ECF);  Roskies, Ralph (EIF);  Ryan, Dan (W);  Rynge, Mats (W);  Sandve, Geir (C);  Santana-Perez, Idafen (W);  Satyanarayanan, Mahadev (W);  Seidel, Ed (E);  Shi, Justin (EIWF);  Silva, Rafael F. (EIWF);  Singer-Villalobos, Faith (E);  Skinner, David (C);  St. Clair, Gloriana (W);  Stodden, Victoria (EIWCF);  Suriarachchi, Isuru (E);  Webster, Keith (W);  Wilkins-Diehr, Nancy (EIWCF)

Key: (E): Event participant; (I): Invited presentation; (W): Written submission; (C): Committee; (F): Final Report contribution